\documentclass{aa}

\usepackage{graphicx}

\usepackage{graphicx,epsfig}
\def\arcsec{$^{\prime\prime}$}
\def\arcmin{$^{\prime}$}
\def\degrees{$^{\circ}$}
\def\etal{\rm et al.~}

\begin{document}

\title{Diffuse radio emission in a REFLEX cluster}
\author{L. Feretti$^{1}$, P. Schuecker$^2$, H. B\"ohringer$^2$, 
F. Govoni$^{1,3}$, G. Giovannini$^{1,3}$}

\offprints{}

\institute{
Istituto di Radioastronomia -- INAF, via Gobetti 101, I--40129
Bologna, Italy
\and 
Max Planck Institut f\"ur Extraterrestrische Physik, PO Box 1312, 
D--85741 Garching, FRG
\and 
Dipartimento di Astronomia, Univ. Bologna, Via Ranzani 1, I--40127 Bologna, 
Italy}

\date{Received ; accepted}

\abstract{ Deep Very Large Array radio observations are presented for
the REFLEX clusters $RXCJ0437.1+0043$ and $RXCJ1314.4-2515$. They are
at similar distance and show similar X-ray luminosity, but they are
quite different in X-ray structure.  Indeed $RXCJ0437.1+0043$ is
regular and relaxed, whereas $RXCJ1314.4-2515$ is characterized by
substructure and possible merging processes.  The radio images reveal
no diffuse emission in $RXCJ0437.1+0043$, and a complex diffuse
structure in $RXCJ1314.4-2515$. The diffuse source in the latter
cluster consists of a central radio halo which extends to the West
toward the cluster periphery and bends to the North to form a possible
relic.  Another extended source is detected in the eastern cluster
peripheral region. Although there could be
plausible optical identifications
for this source, it might also be a relic candidate owing to its very
steep spectrum.  The present results confirm the tight link between
diffuse cluster radio sources and cluster merger processes.
\keywords{ Galaxies: clusters : general -- intergalactic medium --
Radio continuum: general -- X-rays: general } }

\authorrunning{Feretti et al.}
\titlerunning{Diffuse radio emission in a REFLEX cluster}

\maketitle

\section{Introduction}

Some clusters of galaxies contain extended diffuse radio emission with
no optical counterpart with typical sizes of $\sim$ 1 Mpc, low surface
brightness and steep radio spectrum.  These radio sources are
classified as radio halos or relics depending on their central or
peripheral position in the cluster, respectively.  Radio halos and
relics are  associated with the existence of cluster merger processes
(e.g., Buote 2001, Schuecker \etal 2001,
Feretti 2003, Govoni et al. 2004), which provide
the energy to the reacceleration of the radiating particles.  
In particular, radio halos originate from particle reaccelerated in 
the cluster central region, 
possibly by merger turbulence (Fujita et al. 2003,
Brunetti et al. 2004), 
whereas radio relics are tracers of shock waves produced in the
ICM by the flows of cosmological large-scale structure formation
(En{\ss}lin et al. 1998, En{\ss}lin \&  Gopal-Krishna 2001).

The structure of a radio halo shows close similarity to the X-ray
cluster structure (e.g. Deiss et al. 1997; Liang et al. 2000).
Moreover in a number of well-resolved clusters, a spatial correlation
on the large scale between the radio halo brightness and the X-ray
brightness is observed (Govoni et al. 2001a), indicating a connection
between the non-thermal electrons and the thermal intracluster gas.
This correlation is visible e.g. in A2744 also in the {\it Chandra}
high resolution data (Kempner \& David 2004).

The percentage of clusters showing radio halos and relics in the complete
X-ray flux limited sample extracted from Ebeling et al.  (1996) is
$\simeq$ 11\%. The detection rate increases with the X-ray luminosity
 up to $\simeq$ 30\% for the most X-ray luminous clusters
(Giovannini \& Feretti 2002).  The most powerful radio halos and
relics are detected
in the clusters with the highest X-ray luminosity and the largest
total mass, as derived from the radio power-X-ray luminosity and radio
power-cluster mass correlations (Liang \etal 2000, 
Bacchi \etal 2003, Govoni et al. 2001b, Giovannini \& Feretti 2004).
The connection between radio halos and cluster mergers is confirmed by
the radio spectra in A665 and A2163, which are flatter in the regions
influenced by merger processes (Feretti et al. 2004).

According to the previous findings, we expect radio halos and relics
to be common
in X-ray luminous clusters, with irregular X-ray morphology.
Schuecker \& B\"ohringer (1999) showed that the ROSAT All Sky Survey
(RASS) might be helpful to find good cluster candidates housing radio
halos.  The list of clusters from the RASS with 
X-ray luminosity $L_X > 10^{45}$ erg s$^{-1}$, computed assuming
$H_0$ = 50 km s$^{-1}$Mpc$^{-1}$ and q$_0$ = 0.5,  
provides therefore a list of promising clusters hosting a
radio halo, detectable if observed with enough sensitivity.

\begin{figure*}
\includegraphics{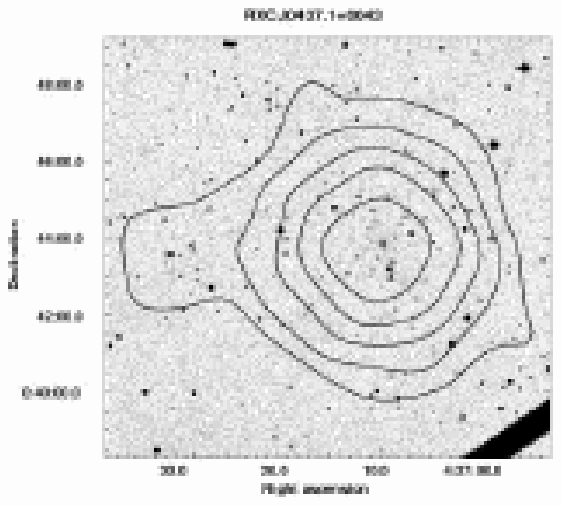}
\includegraphics{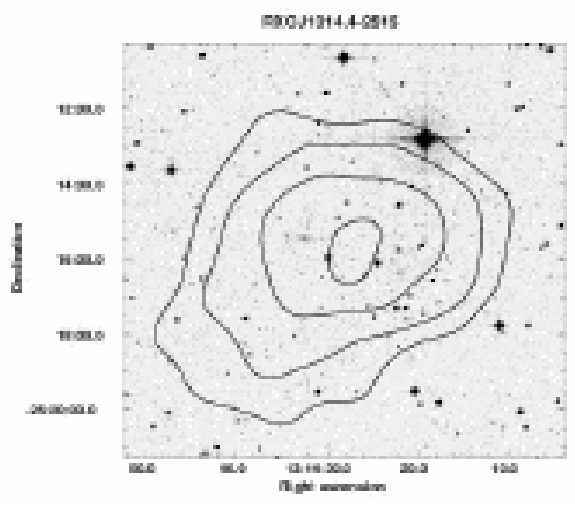}
\vspace{10 cm}
\caption[]{
Contour X-ray images from ROSAT All Sky Survey in the 0.5 - 2 keV 
energy range  of the two clusters 
$RXCJ0437.1+0043$ (left panel) and $RXCJ1314.4-2515$ (right panel),
overlapped onto the grey scale  optical images from the Digital Sky Survey.}
\label{rosat}
\end{figure*}

A well selected sample of X-ray cluster sources found in the southern
ROSAT All-Sky Survey was obtained as a result of the REFLEX cluster
survey (B\"ohringer et al. 2001, 2004).  We inspected the
images of the NRAO VLA Sky Survey (NVSS; Condon et al 1998) for the 20
most X-ray luminous clusters in the overall REFLEX catalogue of 447
clusters and found hints of diffuse radio emission for
$RXCJ0437.1+0043$ and $RXCJ1314.4-2515$, in addition
to  some well known clusters
belonging also to the Abell catalogue.

The cluster $RXCJ0437.1+0043$~ is at redshift 0.2842 and has an X-ray
luminosity of 8.99$\times$10$^{44}$ erg s$^{-1}$ in the 0.1-2.4 keV
band\footnote{Throughout the paper we use the $\Lambda$CDM cosmology with
$H_0$ = 70 km s$^{-1}$Mpc$^{-1}$, $\Omega_m$ = 0.3, and 
$\Omega_{\Lambda}$ = 0.7}.  
The cluster $RXCJ1314.4-2515$~ is at redshift 
0.2439 and has an  X-ray luminosity of 1.09$\times$10$^{45}$ 
erg s$^{-1}$ in the 0.1-2.4 keV band.
These two clusters 
are presented in Fig. \ref{rosat}, where
the low resolution X-ray maps from the ROSAT All Sky Survey, are
overlayed on the grey scale optical images
taken from the Digital Sky Survey.

We obtained sensitive radio observations at 1.4 GHz with
the Very Large Array (VLA), with the aim of searching for diffuse
radio sources in these clusters. 
To properly map these diffuse sources high sensitivity for extended
low brightness feature is needed, but also a good
resolution is necessary to distinguish a real diffuse source from the 
blend of unrelated sources.
 
The similarity between the radio and the X-ray morphology, the high
cluster X-ray temperature and luminosity, and the evidence of
substructure in clusters with radio halos indicate the importance to
consider also the cluster X-ray properties.
Here we present new radio images of these clusters, and
compare them with the X-ray emission. 

In Sect. 2 we present the radio data reduction. In Sects. 3 and 4 we
give the results for the two clusters $RXCJ0437.1+0043$ and
$RXCJ1314.4-2515$, respectively. The discussion and conclusions are
reported in Sect. 5.

With the adopted cosmology, the angular size of 1 arcsec corresponds
to 4.29 kpc at the distance of $RXCJ0437.1+0043$ and to 3.84 kpc at
the distance of $RXCJ1314.4-2515$.

\section{Radio Data}

\begin{table*}
\caption{Details of the VLA observations}
\begin{flushleft}
\begin{tabular}{cccccccc}
\hline
\noalign{\smallskip}
Name &  Frequency  & Bandw.& Config. & Obs. time & Date  \\
RXCJ     & MHz  &   MHz        &         &  hours    &      \\
\noalign{\smallskip}
\hline
\noalign{\smallskip}
$0437.1+0043$ & 1365/1465    & 50   & C & 2.4 &  Jun 2000     \\
                & 1365/1465    & 50   & D & 2.4 &  Aug 2000     \\

$1314.4-2515$ & 1365/1465    & 50   & CnB& 2.4 &  Mar 2000    \\
                 & 1365/1465    & 50   & DnC& 2.4 &  Jul 2000  \\
\noalign{\smallskip}
\hline
\multicolumn{7}{l}{\scriptsize Col. 1: cluster name; Col. 2: observing frequency; Col. 3: bandwidth; Col. 4: VLA configuration;}\\ 
\multicolumn{7}{l}{\scriptsize Col. 5: observing time; Col. 6: observing date.  }\\
\end{tabular}
\end{flushleft}
\label{observ}
\end{table*}

Deep radio observations were obtained for the two clusters with the
more compact configurations of the VLA  
as presented in Table \ref{observ}.  
In order to detect low brightness emission, a good sampling of short
spacings is necessary.  The shortest baseline in the observations is of
35m, corresponding to 200$\lambda$.  This ensures that a large scale
structure of size up to $\sim$ 7.5\arcmin~ is properly imaged.  On the
other hand, the relatively good angular resolution supplied by the
longest baselines, is important to separate discrete radio sources.
We note that given the southern declination of $RXCJ1314.4-2515$, the
VLA was used in hybrid configurations (CnB and DnC) in order to obtain
a roughly circular restoring beam. 

 For the cluster $RXCJ0437.1+0043$, the sources 3C48 and J0423-013 were
observed as calibrators of the flux density scale, and of the antenna
gains and phases, respectively.  For $RXCJ1314.4-2515$, the source
3C286 was observed for the flux density scale calibration, and the
source J1284-199 was observed as gain and phase calibrator.  

The data were calibrated and reduced with the Astronomical Image
Processing System (AIPS), following the standard procedure.
Significant editing of the uv data was needed to identify and
remove bad data.
Images were produced by Fourier-Transform, Clean and Restore,
using the task IMAGR with uniform uv weight,
no Gaussian taper, and ROBUST = 0. Several
cycles of imaging and self-calibration were performed, to minimize the
effects of amplitude and phase variations. At each step,
clean components were carefully selected 
as a model for the self-calibration.
Gain calibration solutions were obtained  with a long integration time,
after a few cycles of phase calibration only.
The process was stopped when no further significant improvement was
obtained. Then, a deep cleaning was applied to the data,
until the total cleaned flux reached a stable value. 

Data from the two different configurations were reduced separately, in
order to analyse the possible existence of spurious features.
Images at intermediate resolutions were also obtained,
by adding together the data from the two configurations. 

The achieved sensitivities in the low-resolution maps are 0.06 mJy/beam
$RXCJ0437.1+0043$ and 0.068 mJy/beam in $RXCJ1314.4-2515$. 
The confusion level expected in the VLA images in the D configuration
is $\sim$ 70 $\mu$Jy/beam (J. Condon and R. Perley, private communication), 
thus in our images we reach the confusion limit.

The calibration of the polarized emission was not possible because of
technical problems.  Therefore, polarization information cannot be
derived from the present radio data.

\begin{table*}
\caption{Radio sources at the center of $RXCJ0437.1+0043$}
\begin{flushleft}
\begin{tabular}{cccccccr}
\cline{1-5}
\noalign{\smallskip}
 Radio  & & Optical & &   \\
RA (2000) Dec &  S$_{1.4~GHz}$ &  RA (2000) Dec &  m$_B$ \\
 h ~ m ~ s ~~~ \degrees  ~ \arcmin ~ \arcsec & mJy &  
h ~ m ~ s ~~~ \degrees  ~ \arcmin ~ \arcsec &    \\
\noalign{\smallskip}
\cline{1-5}
\noalign{\smallskip}
04 37 07.8 +00 44 23 & 11.6 $\pm$ 0.2 & 04 37 07.88 +00 44 22.8 & 21.0 \\
04 37 09.6 +00 43 55  & 1.1 $\pm$ 0.1 & 04 37 09.49 +00 43 51.1 & 18.9 \\
\noalign{\smallskip}
\cline{1-5}
\multicolumn{8}{l}{\scriptsize Col. 1: radio source position; 
Col. 2: flux density at 1.4 GHz;}\\ 
\multicolumn{8}{l}{\scriptsize Col. 3: optical position from superCOSMOS; 
Col. 4: blue magnitude.
 }\\
\end{tabular}
\end{flushleft}
\label{observ}
\end{table*}

\section{Results on RXCJ0437.1+0043}

\begin{figure}
\includegraphics{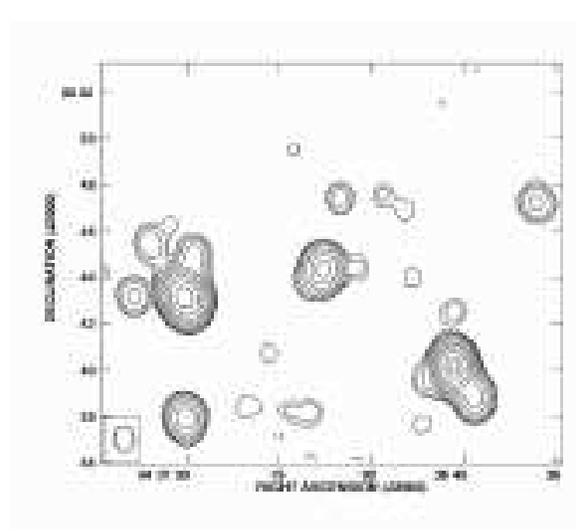}
\vspace{9 cm}
\caption[]{Contour radio image of the cluster $RXCJ0437.1+0043$
obtained with the D array at 1.4 GHz, with resolution of 
63.9\arcsec$\times$48.9\arcsec (FWHM at PA= 5\degrees). The rms noise 
level is 0.06 mJy/beam. Contour levels
are -0.2, 0.2, 0.3, 0.5, 1, 2, 4, 8, 16, 32, 48 mJy/beam.
In this image, and in the following images, the observing beam is
indicated at the bottom left corner.
}
\label{r0437d}
\end{figure}

\begin{figure}
\includegraphics{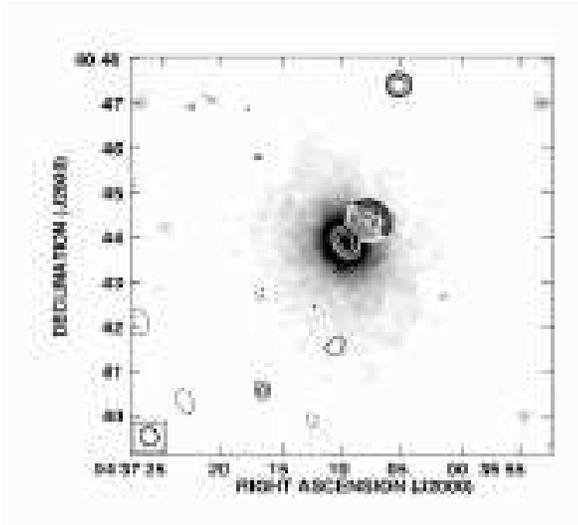}
\vspace{9 cm}
\caption[]{Contour radio image of the cluster $RXCJ0437.1+0043$
obtained at 1.4 GHz with the C array, 
overlayed on the grey scale X-ray emission
detected by XMM.
The FWHM is 23.7\arcsec$\times$20.9\arcsec (at PA= 23\degrees). The rms noise 
level is 0.06 mJy/beam. Contour levels
are -0.2, 0.2, 0.3, 0.5, 1, 2, 4, 8 mJy/beam.
}
\label{rx0437}
\end{figure}

\begin{figure}
\includegraphics{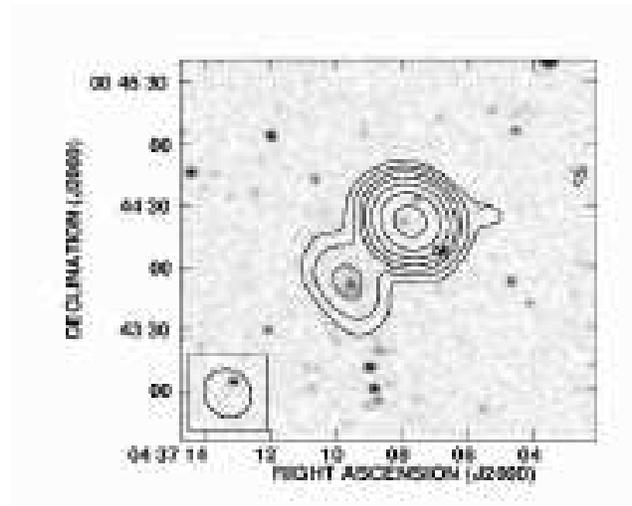}
\vspace{9 cm}
\caption[]{Contour radio image of the cluster $RXCJ0437.1+0043$
from Fig. \ref{rx0437}, superposed on the grey-scale optical image 
from the DSS2 red filter. Contour levels are
 0.15, 0.3, 0.6, 1, 2, 4, 8 mJy/beam.
}
\label{ro0437}
\end{figure}

This cluster belongs to the REFLEX-DXL sample defined by Zhang et
al. (2004).  In the X-ray image, obtained with XMM in the 0.5-2 keV
energy range (B\"ohringer et al. 2005, in preparation), the cluster
looks relatively symmetric with some elongation.  It has a very bright
center, with the temperature falling from about 7-8 keV in the outer
region to about 5 keV in the center.  Thus it is likely a relaxed
cluster, showing a cooling core.

The radio emission of the cluster, as detected at low
resolution, is presented in Fig \ref{r0437d}.  The field shown in the
figure covers a much larger cluster area than that shown in the left panel
of Fig. \ref{rosat}.  There are several discrete radio sources, and no
evidence of any large scale diffuse emission. 
The higher resolution image of the central cluster region, 
obtained with the C array, is shown in
Fig. \ref{rx0437}, overlayed on the grey scale X-ray emission.  The
only radio emission at the center of this cluster is represented by
two close unresolved radio sources, of 1.1 and 11.6 mJy respectively,
one coincident with the X-ray
peak, the other slightly displaced from it (see Tab. 2). Both sources
are likely identified with galaxies (see Fig. \ref{ro0437} and
Tab. 2).

No diffuse radio emission is detected in this cluster, at the
brightness level of 0.18 mJy/beam (3$\sigma$), corresponding to $\sim$
0.06 $\mu$Jy arcsec$^{-2}$.  This value is lower than
the average brightness of the radio halo Coma C in the Coma cluster.
The maximum size of an extended feature
detectable with the present observations is of
$\sim$ 7.5\arcmin, which corresponds to a linear size of $\sim$ 1.9 Mpc.
Thus a radio halo with a regular structure, similar in brightness
to Coma C, and extended  $<$ 1.9 Mpc
would have been easily detected by the present observations.

\section{Results on  RXCJ1314.4$-$2515}

\begin{table*}
\caption{Extended radio sources in  $RXCJ1314.4-2515$}
\begin{flushleft}
\begin{tabular}{cccccccc}
\hline
\noalign{\smallskip}
 Radio  & & & Optical \\
Source  &  S$_{1.4~GHz}$  &  S$_{1.4~GHz}$   &
 RA (2000) Dec &  m$_B$  \\
  &  DnC arr  &   CnB arr  \\
& mJy &  mJy & h ~ m ~ s ~~~ \degrees  ~ \arcmin ~ \arcsec      \\
\noalign{\smallskip}
\hline
\noalign{\smallskip}
H + R1 &  40.5 $\pm$ 0.6 & & \\
&\\
R1 &  & 20.2 $\pm$ 0.5 & 13 14 19.60 $-$25 15 38.0 & 16.1 \\
   &        &         & 13 14 19.30 $-$25 15 30.0 & 20.1 \\
   &        &         & 13 14 16.86 $-$25 15 58.9  & 18.1 \\
&\\
R2 & 11.1 $\pm$ 0.3  & 10.1 $\pm$ 0.3 & 13 14 44.54 $-$25 14 36.7 & 18.2  \\
   &        &         &  13 14 46.19 $-$25 15 09.0 & 21.4 \\
\noalign{\smallskip}
\hline
\multicolumn{7}{l}{\scriptsize Col. 1: source name; Col. 2: 
1.4 GHz flux density at lower resolution;}\\
\multicolumn{7}{l}{\scriptsize Col. 3: 
1.4 GHz flux density at higher resolution;}\\
\multicolumn{7}{l}{\scriptsize Col. 4: position of optical 
objects from superCOSMOS; Col. 5: blue magnitude  }\\
\end{tabular}
\end{flushleft}
\label{tab3}
\end{table*}

\subsection {Large scale cluster radio emission}

\begin{figure}
\includegraphics{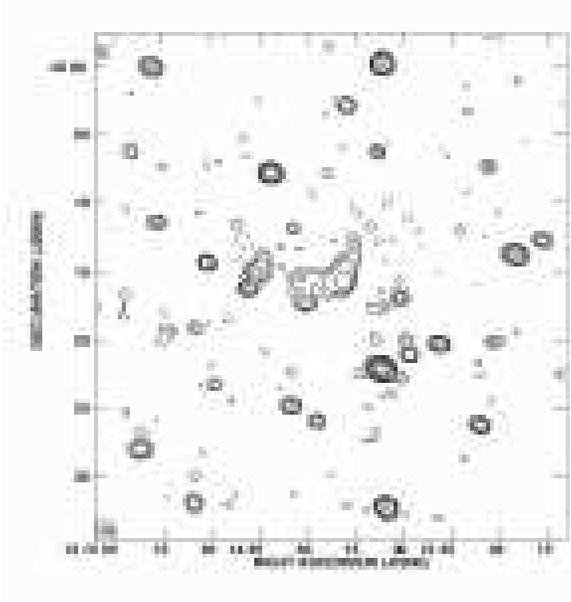}
\vspace{9 cm}
\caption[]{Radio image of the cluster $RXCJ1314.4-2515$
obtained at 1.4 GHz with the DnC array at the 
 resolution of 41.8\arcsec$\times$34.0\arcsec 
(at PA= 75\degrees). The rms noise level is 0.068 mJy/beam.
Contour levels are -0.5, -0.2, 0.2, 0.5, 1.0, 
3.0, 10.0, 30.0, 80.0 mJy/beam.}
\label{r1314d}
\end{figure}

The radio image presented in Fig. \ref{r1314d} has been obtained
with the data of the DnC array with a deep cleaning, down to 0.07
mJy/beam, which corresponds approximately to the map noise level.
The figure shows a large field, down to about
50\% of the antenna primary beam, with the lowest positive contour
corresponding to 3 times  the noise level.
Diffuse emission is easily visible at the cluster center,
with a brightness of about 10 times the noise level.
No other either negative or positive features are present
in the map at the same level, ensuring that the diffuse
emission cannot be an artifact of the data reduction process.

\begin{figure}
\includegraphics{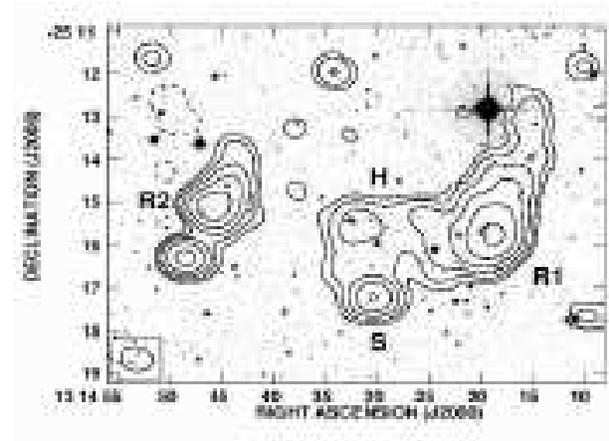}
\vspace{9 cm}
\caption[]{Contour radio emission of the central region of 
 $RXCJ1314.4-2515$ overlayed on the grey-scale red image from the
DSS2. The radio image, obtained from the DnC array data,  
has a resolution of 41.8\arcsec$\times$34.0\arcsec~(FWHM at PA = 75\degrees).
and a  noise level of  0.068 mJy/beam.
Contour levels are  -0.2, 0.2, 0.4, 0.8, 1.6, 3.2, 6.4  mJy/beam.
Relevant regions are  marked by labels (see text).
}
\label{ro1314}
\end{figure}

Fig. \ref{ro1314} shows the enlargement of the central cluster 
region, with the radio emission overlayed on the optical image.

\begin{figure}
\includegraphics{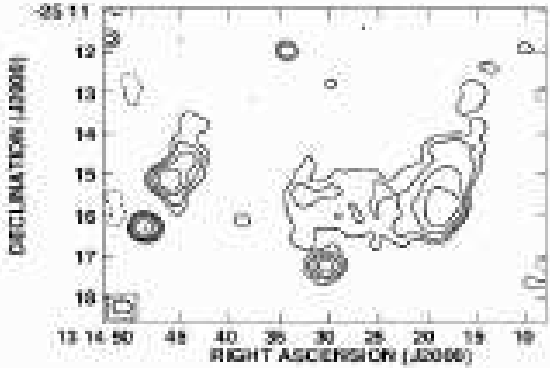}
\vspace{8 cm}
\caption[]{Radio image of the cluster $RXCJ1314.4-2515$
at 1.4 GHz obtained with the combined CnB and DnC array,
with the resolution of 24.4\arcsec$\times$20.4\arcsec (at 
PA= 77.9\degrees)
The rms noise level is 0.09 mJy/beam. Contour levels are
-0.27, 0.27, 0.5, 1, 2, 4, 8  mJy/beam.
}
\label{r1314ir}
\end{figure}

In Fig.  \ref{r1314ir} we present an image at
intermediate resolution obtained from the combination of the data from
the DnC and CnB arrays, using UVTAPER=10  and ROBUST = 0 with the 
AIPS task IMAGR. A deep cleaning down to about 1 $\sigma$ level was
applied. 
The two images are in very good agreement, confirming the diffuse
nature of the extended emission detected at lower resolution.  The
slightly higher noise level in the intermediate resolution image with
respect to the low resolution image is probably due to the lower
weight of the shortest baselines, where most of the diffuse flux
density is present.

The diffuse radio emission in this cluster is complex.  The
most relevant feature (labeled H in Fig. \ref{ro1314}) permeates the
cluster center, around the brightest cluster galaxy, and is
elongated approximately in E-W direction. Its brightest region is on
the western side (R1) and coincides with a prominent bend to the
North.   The total flux density of this emission is
given in Tab. 3. Its total extent is at least 8\arcmin,
corresponding to $\sim$ 1.8 Mpc. The point source at the southern
boundary of the diffuse feature (S) is an unrelated discrete source
(see the higher resolution map of Fig. \ref{r1314c}).  Another extended source
showing an elongated structure (R2 in Fig. \ref{ro1314}) is detected
in the eastern cluster region.

\subsection {Sources R1 and R2}

\begin{figure}
\includegraphics{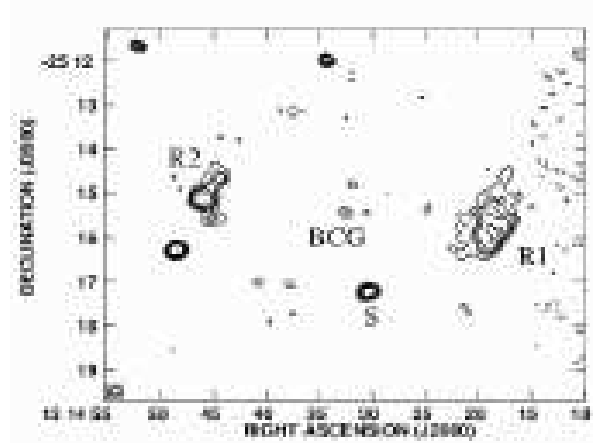}
\vspace{8 cm}
\caption[]{Radio image of the cluster $RXCJ1314.4-2515$
at 1.4 GHz obtained with the CnB array
with the resolution of 12.8\arcsec$\times$9.3\arcsec (at 
PA= 88.5\degrees)
The rms noise level is 0.045 mJy/beam. Contour levels are
-0.15, 0.15, 0.3, 0.4 ,0.6, 1, 3, 6, 10, 20, 50, 80 mJy/beam.
}
\label{r1314c}
\end{figure}

The high-resolution image of the cluster central region, 
 obtained with the CnB array data, is
shown in Fig.  \ref{r1314c}. Most of the extended structure is
resolved out, confirming that it is really diffuse and cannot be due
to discrete sources.  The only discrete source detected within
the central region of the diffuse emission is very faint, 
with a total flux of $\sim$ 0.5 mJy, and is identified with
the brightest cluster galaxy (BCG in Fig. \ref{r1314c}), in agreement
with Valtchanov et al. (2002). No jet-like structure is visible.

\begin{figure}
\includegraphics{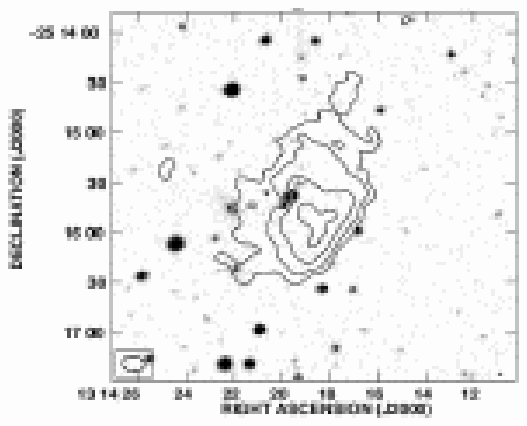}
\vspace{9 cm}
\caption[]{Enlargement of the region R1 of Fig.
 \ref{r1314c}, overlayed on the grey-scale optical image taken
from the DSS2 red filter. Contour levels are  0.15, 0.3, 0.6, 1  mJy/beam.
}
\label{ro1314w}
\end{figure}

\begin{figure}
\includegraphics{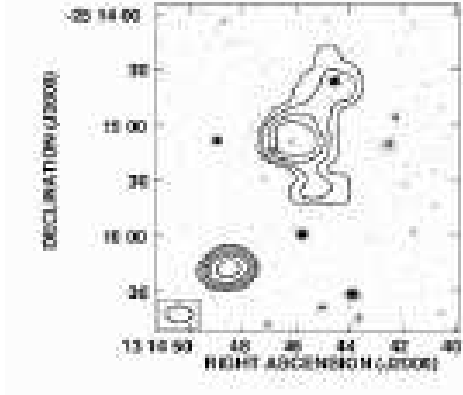}
\vspace{9 cm}
\caption[]{Enlargement of the region R2 of Fig.
 \ref{r1314c}, overlayed on the grey-scale optical image taken
from the DSS2 red filter.
Contour levels are  0.15, 0.3, 0.6, 1, 2  mJy/beam.
}
\label{ro1314e}
\end{figure}

The region of highest brightness R1 is detected as a feature with
amorphous morphology and a steep brightness gradient on the S-W
edge. This source accounts for about 50\% of the flux density of the
whole diffuse emission detected at low resolution (Tab. 3). 
The diffuse source R2 is resolved in at least 3 components, with an
overall structure reminiscent of that of a wide-angle tailed (WAT) radio
source, however its central component is likely extended.

A radio image of this cluster, obtained at 1.4 GHz with the Australia
Telescope Compact Array (ATCA) in the 6C configuration with a resolution
of 23.8\arcsec $\times$ 9.4\arcsec~ is presented by Valtchanov et
al. (2002). This image is very similar to our high resolution map of
Fig.  \ref{r1314c}.  Also in the ATCA map, as in Fig.  \ref{r1314c},
the lower brightness emission is missing. This is consistent
with the fact that this emission is diffuse, thus a more compact ATCA
configuration than the 6C configuration would be needed to detect it.

Valtchanov et al. (2002) identify the faint radio source at the center
with the brightest cluster galaxy (BCG) and discuss the possibility
that R1 and R2 are the symmetric lobes of an FRII radio galaxy with
the core coincident with the BCG. However, they consider this
interpretation quite questionable,  as the radio galaxy is not located
in an underdense environment as expected for such a giant 
radio galaxy.  

We have searched if there are plausible identifications for sources R1
and R2.  The positions and magnitudes of the optical candidates 
are given in Tab. 3.  From the radio -
optical overlay presented in Fig. \ref{ro1314w}, it turns out that the
source R1 is not obviously associated with any optical galaxy. 
Among the objects listed in Tab. 3, i.e. those well within the radio 
contours, the brightest object is a star (Valtchanov et al. 2002).

The radio - optical overlay of the source R2 (Fig. \ref{ro1314e})
shows that two optical objects could be responsible for the radio
emission (see Tab. 3): a faint galaxy coincident with the northern
component, and a fainter galaxy located approximately at the peak of
the middle component. In the first case, the source would be a tailed
radio galaxy, but showing an unusual structure with a very bright
region at the center of the tail.  In the second case, the source
would be a WAT radio galaxy, but this seems unlikely, since WAT
sources are usually located at the cluster centers.  Alternatively, if
the radio emission would be due to both optical objects, the northern
radio galaxy would be slightly extended, the other radio galaxy,
consisting of the central and southern components of R2, would show a
tailed structure.  The spectral information, available from Valtchanov
et al. (2002), does not seem to support the identification of the
source R2 with radio galaxies.  Indeed, those authors derive a very
steep spectrum with $\alpha^{2.5}_{1.4}$ $\sim$ 2.5 (S$_{\nu}$
$\propto$ $\nu^{-\alpha}$), and suggest that this source is an
excellent candidate for a relic source.  

\subsection {Radio - X-ray comparison}

\begin{figure}
\includegraphics{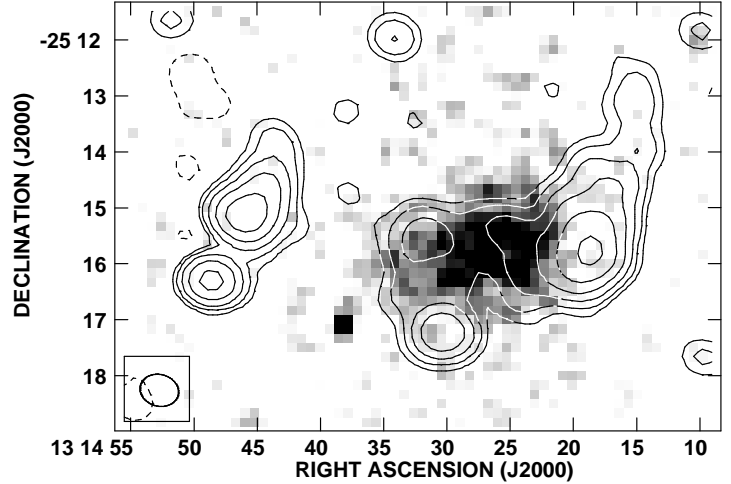}
\vspace{9 cm}
\caption[]{Contour radio image of the cluster $RXCJ1314.4-2515$
at 1.4 GHz, superposed on the grey scale X-ray emission obtained
by ROSAT HRI.
Contour levels are -0.2, 0.2, 0.4, 0.8, 1.6, 3.2, 6.4
mJy/beam.
}
\label{rx1314}
\end{figure}

The large scale X-ray emission of this cluster (Fig. \ref{rosat},
right panel) shows an asymmetric brightness distribution, with
substructure to the S-E, which is indication of a recent cluster
merger. An ongoing merger at the cluster center is also derived by the
ROSAT HRI map (Valtchanov et al.  2002, and Fig.\ref{rx1314}), which
shows emission clearly elongated approximately in the E-W direction.
From the overlay between the radio and X-ray HRI emission
(Fig.\ref{rx1314}), it is evident that the diffuse radio emission
permeates the cluster center and extends toward the West to the
cluster periphery, then bending to the North.

 On the basis of the typical
classification of radio halos and relics in clusters,
the diffuse radio emission at the cluster center, coincident with the
region of highest X-ray brightness, can be classified as a radio halo,
with irregular and elongated morphology.
The diffuse structure including source R1 and extending toward the North 
could be classified as a relic.  This structure 
has a total size of $\sim$ 4\arcmin, i.e. about 900 kpc.
It is located at $\sim$ 3.5\arcmin~ from the cluster center, i.e. at a
projected distance of about 800 kpc.
We note that this peripheral diffuse emission is significantly more
extended toward the North than the region R1 imaged at higher
resolution and classified as a relic by Valtchanov et al. (2002).

On the other hand, since the diffuse emission looks like a single
radio feature, we could consider it as a very unusual and peculiar
cluster radio halo.  From the total flux density at 1.4 GHz of the
extended feature H, including the peripheral structure (Tab. 3), a
monochromatic radio power of 7.3 $\times$ 10$^{24}$ W Hz$^{-1}$ is
obtained for this source.
This is about 10 times higher than the power of the radio halo in
Coma, and is comparable to that of the most powerful halos in hot
clusters.  Thus, on the basis of the power, we cannot exclude the
possibility that the whole feature is a very unusual radio halo. More
data are necessay to clarify this point and check if this source is
consistent with the radio power- X-ray luminosity correlation (Bacchi
et al. 2003, Govoni et al. 2001b).

The western source R2 is located in a peripheral cluster region 
at a projected distance of $\sim$ 800 kpc from the cluster center
and  has a total  extent of about 500 kpc,
with the major axis roughly perpendicular to 
the cluster radius. Valtchanov et al. (2002) suggested that it
is a candidate for a relic source, owing to its extent, location 
and very steep spectrum.
From our data, it is not possible to firmly establish its nature.
We find that it may have plausible optical identifications,
however more detailed information on its structure and spectrum
are needed to understand if this emission is due to a relic or
to discrete radio galaxies.


\section{Discussion and conclusions}

The two clusters studied here have similar redshift and X-ray
luminosity, but show different X-ray structure.  The cluster
$RXCJ0437.1+0043$ is rather regular and relaxed, and is characterized
by a cooling core, whereas the cluster $RXCJ1314.4-2515$ shows
substructure to the S-E on the large scale, and elongation in the E-W
direction in the innermost region, thus it shows merging activity.

Diffuse emission on a scale of $\sim$ 1.8 Mpc is detected in
$RXCJ1314.4-2515$. This emission is rather irregular in shape,
extending to the peripheral western cluster boundary and bending to
the North.  We suggest that this morphology is likely to consist of a
central radio halo, and a peripheral relic. The two diffuse sources
are strictly connected to each other, making this diffuse emission
quite peculiar. Another case of complex diffuse emission in the same
cluster is in A754 (Kassim et al. 2001, Bacchi et al. 2003), where
there is a central halo and a more peripheral relic, which show
atypical structures and may be connected to each other.  

The elongation of the radio structure at the cluster center is in
the same direction of the X-ray brightness distribution detected at
high resolution by the ROSAT HRI. Since the E-W elongation of the
X-ray brightness distribution is likely indicative of a cluster merger
in that direction, the radio structure confirms the existence of a
tight connection between the radio halo and the central merger.  The
source R1 is located along the major axis of the HRI X-ray brightness
distribution, i.e. along the merger axis, where merger shocks are
expected to be present. Its structure is elongated nearly
perpendicularly to the merger axis.  In the framework of models
that suggest relics to trace cluster merger shock waves (En{\ss}lin et
al. 1998, En{\ss}lin \& Gopal-Krishna 2001), both the location and the
structure of R1 seem to be consistent with this source being a cluster
relic.  

In addition to the diffuse emission discussed above, there is also
the extended radio emission R2 in the eastern cluster region, which  may
be a candidate for another relic source (see also Valtchanov et
al. 2002).
We could speculate that the possible existence of a second
cluster relic would be consistent with the dynamical activity derived
in this cluster. Indeed, the two relics would be located at opposite
sides of the cluster, along the merger axis, with their structure
elongated roughly perpendicularly to the merger axis. In this respect,
the cluster $RXCJ1314.4-2515$ would be similar to A3667 (R\"ottgering
et al. 1997), although in the cluster A3667 the two opposite relics
are much more extended and no central halo is detected.

The presence of diffuse radio sources in $RXCJ1314.4-2515$, together
with the lack of a radio halo in $RXCJ0437.1+0043$, confirms that the
formation of diffuse emission is likely associated with the cluster
merger activity.  A consistent, although qualitative, picture is that
$RXCJ1314.4-2515$ exhibits a central halo, linked to the central
merger, and relic emission originating from the peripheral shocks.
Future sensitive and high resolution data both in radio and X-ray are
needed to better investigate the properties of the cluster
$RXCJ1314.4-2515$, and to study the details of the connection between
merger processes and the formation of halos and relics.

\section*{Acknowledgments}

We thank the anonymous referee for his/her suggestions, which
have significantly improved the presentations of the results.
NRAO is a facility of the National Science Foundation, operated under
cooperative agreement by Associated Universities, Inc.  This work was
partly supported by the Italian Space Agency (ASI).

\end{document}